\documentclass[%
reprint,
superscriptaddress,
 amsmath,amssymb,
 aps,
prb,
]{revtex4-1}

\usepackage{graphicx}
\usepackage{dcolumn}
\usepackage{bm}

\usepackage{siunitx}
\usepackage{upgreek}

\begin{document}

\preprint{APS/123-QED}

\title{Optical properties and interparticle coupling of plasmonic bowtie nanoantennas on a semiconducting substrate}

\author{K. Schraml}
\email{konrad.schraml@wsi.tum.de}
\affiliation{Walter Schottky Institut and Physik Department, Technische Universit\"at M\"unchen, Am Coulombwall 4, 85748 Garching, Germany\\Nanosystems Initiative Munich, Schellingstra{\ss}e 4, 80799 M\"unchen, Germany}
\homepage{http://www.wsi.tum.de}
\author{M. Spiegl}
\affiliation{Walter Schottky Institut and Physik Department, Technische Universit\"at M\"unchen, Am Coulombwall 4, 85748 Garching, Germany\\Nanosystems Initiative Munich, Schellingstra{\ss}e 4, 80799 M\"unchen, Germany}
\homepage{http://www.wsi.tum.de}
\author{M. Kammerlocher}
\affiliation{Walter Schottky Institut and Physik Department, Technische Universit\"at M\"unchen, Am Coulombwall 4, 85748 Garching, Germany\\Nanosystems Initiative Munich, Schellingstra{\ss}e 4, 80799 M\"unchen, Germany}
\homepage{http://www.wsi.tum.de}
\author{G. Bracher}
\affiliation{Walter Schottky Institut and Physik Department, Technische Universit\"at M\"unchen, Am Coulombwall 4, 85748 Garching, Germany\\Nanosystems Initiative Munich, Schellingstra{\ss}e 4, 80799 M\"unchen, Germany}
\homepage{http://www.wsi.tum.de}
\author{J. Bartl}
\affiliation{Walter Schottky Institut and Physik Department, Technische Universit\"at M\"unchen, Am Coulombwall 4, 85748 Garching, Germany\\Nanosystems Initiative Munich, Schellingstra{\ss}e 4, 80799 M\"unchen, Germany}
\homepage{http://www.wsi.tum.de}
\author{T. Campbell}
\affiliation{Institute for Critical Technology and Applied Science (ICTAS), Virginia Polytechnic Institute and State University (Virginia Tech), Blacksburg, VA 24061, USA}
\author{J. J. Finley}
\affiliation{Walter Schottky Institut and Physik Department, Technische Universit\"at M\"unchen, Am Coulombwall 4, 85748 Garching, Germany\\Nanosystems Initiative Munich, Schellingstra{\ss}e 4, 80799 M\"unchen, Germany}
\homepage{http://www.wsi.tum.de}
\author{M. Kaniber}
\affiliation{Walter Schottky Institut and Physik Department, Technische Universit\"at M\"unchen, Am Coulombwall 4, 85748 Garching, Germany\\Nanosystems Initiative Munich, Schellingstra{\ss}e 4, 80799 M\"unchen, Germany}%
\homepage{http://www.wsi.tum.de}



\date{\today}

\begin{abstract}
We present the simulation, fabrication and optical characterization of plasmonic gold bowtie nanoantennas on a semiconducting GaAs substrate as geometrical parameters such as size, feed gap, height and polarization of the incident light are varied. The surface plasmon resonance was probed using white light reflectivity on an array of nominally identical, 35\,nm thick Au antennas. To elucidate the influence of the semiconducting, high refractive index substrate, all experiments were compared using nominally identical structures on glass. Besides a linear shift of the surface plasmon resonance from 1.08\,eV to 1.58\,eV when decreasing the triangle size from 170\,nm to 100\,nm on GaAs, we observed a global redshift by 0.25\,$\pm$\,0.05\,eV with respect to nominally identical structures on glass. By performing polarization resolved measurements and comparing results with finite difference time domain simulations, we determined the near field coupling between the two triangles composing the bowtie antenna to be $\sim$8$\times$ stronger when the antenna is on a glass substrate compared to when it is on a GaAs substrate. The results obtained are of strong relevance for the integration of lithographically defined plasmonic nanoantennas on semiconducting substrates and, therefore, for the development of novel optically active plasmonic-semiconducting nanostructures.

\end{abstract}

\maketitle 


\section{\label{sec:level1}Introduction}
Resonant metallic nanoantennas in the optical regime have generated much interest over the last decade due to their ability to confine light to deep subwavelength dimensions\cite{barnes2003surface,schuller2010plasmonics,novotny2011antennas,gramotnev2014nanofocusing}. In particular, coupled nanoparticle dimers are of interest since they provide a large electric field enhancement within the feed gap\cite{kim2008high,hanke2012tailoring}. This effect holds great promise for new applications in sensing\cite{punj2013plasmonic} and in fluorescence enhancement,\cite{anger2006enhancement,kuhn2006enhancement,pompa2006metal,kinkhabwala2009large,acuna2012fluorescence} as well as for emission control\cite{taminiau2008optical,Curto20082010,curto2013multipolar} of single molecules and quantum emitters. Triangular-shaped nanoparticles in a tip-to-tip configuration, the so-called bowtie nanoantenna, are used to take advantage of the lightning rod effect. The optical response of such plasmonic nanopantennas has been studied as a function of different geometrical parameters such as size\cite{muhlschlegel2005resonant,fischer2008engineering,hanke2009efficient,prangsma2012electrically}, feed gap\cite{fromm2004gap,schuck2005improving,merlein2008nanomechanical,hanke2009efficient} and shape\cite{sonnichsen2002drastic,fischer2008engineering}, materials\cite{wang2006general} and wavelength range\cite{fromm2004gap,rivas2004propagation,hibbins2005experimental}. Beside the widely used glass substrates, plasmonic nanoparticles have also been investigated on semiconductor substrates like Si or GaAs with relation to their use in photovoltaic applications\cite{lim2007photocurrent,catchpole2008plasmonic,atwater2010plasmonics}. Recently, it has been theoretically shown that the use of high refractive index substrates such as semiconductors can boost the radiative decay rate of quantum emitters by a factor of \textgreater 7500 when the optical and geometrical properties of the quantum emitters and plasmonic nanoantenna are properly engineered\cite{chen2012metallodielectric}. 

In this paper, we present a comprehensive study of the optical properties of lithographically defined gold bowtie nanoantennas on GaAs. The results obtained are compared to nominally identical structures on glass to gain deeper insights into the effect of the high refractive index substrate on the plasmonic response in the optical regime. Complementary finite difference time domain (FDTD) simulations were employed to find the optimized thickness (t) of the structures and to compare our experimental findings with predictions based on classical electrodynamics. We optically probed the surface plasmon resonance (SPR) frequency using white light reflectivity as the size (s), feed gap (g) and the polarization of the incident electromagnetic field ($\Theta$) are varied. The SPR shifts linearly from 1.08\,eV to 1.58\,eV when decreasing the triangle size from 170\,nm to 100\,nm on GaAs, similar to the shift measured on glass over the same range. We observe a near uniform redshift of 0.25\,$\pm$\,0.05\,eV upon moving from glass to GaAs substrates. Furthermore, the SPR strongly depends on the feed gap between the two triangles. With decreasing feed gap size we observe a redshift of the SPR that follows a $\text{g}^{-3}$-dependence, indicative of dipole-dipole coupling between the two particles. The absolute shift between g\,=\,80\,nm and g\,=10\,nm was found to be 0.03\,eV on GaAs, much smaller that the 0.20\,eV observed on glass, indicative of a weaker coupling strength due to the presence of the high refractive index substrate. We quantified this interparticle coupling to be $\sim$8$\times$ lower on GaAs compared to a glass substrate when probing the coupled and uncoupled mode of the bowtie using polarization resolved measurements. Our simulations indicate that this effect originates from the strongly modified electric field distribution due to the presence of the high refractive index substrate and the presence of a thin native oxide layer on top of it.

%
%
\section{\label{sec:level2}Fabrication \& Experimental Setup}

\begin{figure}[t!]
\includegraphics[width=\columnwidth]{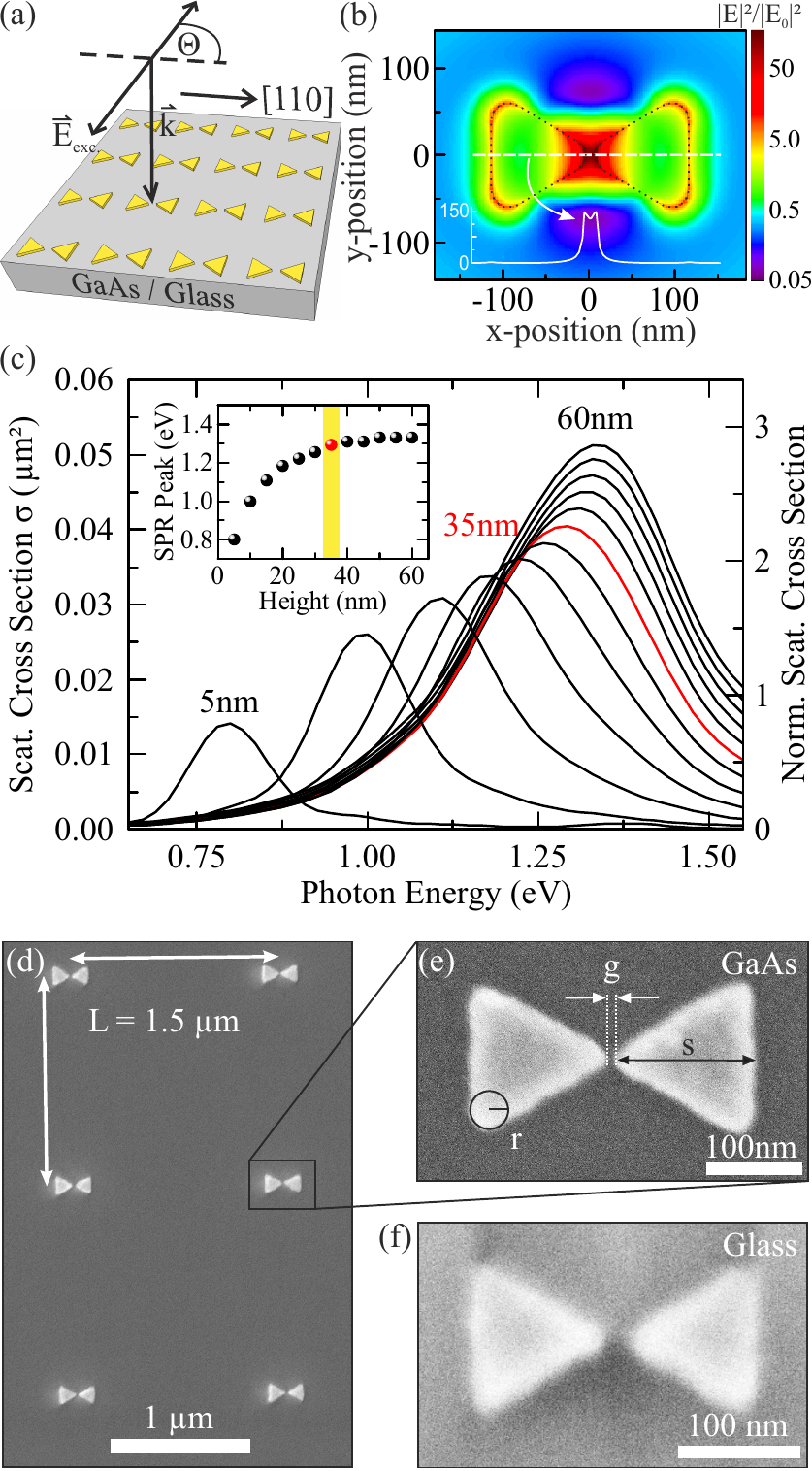}
\caption{(a) Schematic illustration of the sample layout. Bowties are defined in arrays with a pitch of \SI{1.5}{\micro\metre} to avoid near field coupling. Structures are illuminated from top with light polarized along the long axis of the bowtie $\Theta$\,=\,0$^\circ$. (b) FDTD simulation reveals electric field distribution (log-scale) of a single bowtie on GaAs (s\,=\,110\,nm, g\,=\,10\,nm) 2\,nm above the gold surface at the electric dipole resonance. Inset shows the cross section along the dashed line at y\,=\,0\,nm (linear scale). (c) Calculated scattering spectra for varying t. Inset shows SPR peak position as function of t with highlighted optimum t\,=\,35\,nm. The scale on the right hand side is normalized to the geometrical area of the bowtie. (d) SEM image of a fabricated array and a close-up of a single bowtie (e) on GaAs show  gaps, and tip radii close to the resolution limit of our e-beam system with yield of almost 100\,\%. (f) SEM image for similar structures on glass substrates.} 
\label{sample}
\end{figure}

Gold bowtie nanoantennas were defined on undoped GaAs [100] substrates using standard electron beam lithography. As depicted schematically in fig. \ref{sample}(a), they were arranged in arrays to enhance the signal in the white light reflectivity measurements. We chose a pitch of \SI{1.5}{\micro\meter} to avoid near-field coupling between two adjacent bowties\cite{quinten1998electromagnetic,weeber1999plasmon} whilst retaining the possibility to address single bowties with a focused laser beam in future experiments. 

Prior to the fabrication process, we performed FDTD simulations using a commercially available software package (Lumerical Solutions, Inc.\cite{test}) to identify the optimum thickness of the Au-film used to define our nanostructures. For future combination with semiconductor quantum emitters such as InGaAs quantum dots\cite{shields2007semiconductor}, it is highly desirable to overlap the bowtie's surface plasmon resonance with the emission range of the dots at $\sim$\,1.3\,eV while preserving a high electric field enhancement within the feed gap\cite{chen2012metallodielectric}. Fig. \ref{sample}(b) shows a typical result of the simulated electric field enhancement, defined as the ratio of the intensity of the electric field $|E|^2$ 2\,nm above the the bowtie surface compared to the intensity of the incoming plane wave $|E_0|^2$. The simulation was performed for a single bowtie (s\,=\,110\,nm, g\,=\,10\,nm) and probed at the electric dipole resonance $\text{E}_{\text{res}}$\,=\,1.33\,eV. Here, we observe that the E-field is mostly concentrated in an area of $\sim$30$\times$30\,nm\textsuperscript{2} with electric field enhancements up to a factor of $|E|^2/|E_0|^2$\,=\,180$\times$ using the definition specified above. The inset shows the cross section of the field amplitude along the dashed line at y\,=\,0\,nm on a linear scale. We calculated that 79\% of the intensity along this curve is concentrated within a region of size -15\,nm\,$\leq$\,x\,$\leq$\,15\,nm. Hence, the bowtie geometry is capable of focusing light into spatial regions, similar to the lateral dimension of a single self-assembled InGaAs quantum dot\cite{marquez2001atomically}. Besides the electric field distribution, we also calculated the scattering cross section $\upsigma$ of the nanoantenna, which is defined as P\,=\,$\upsigma\cdot\text{I}_{\text{0}}$, where $\text{I}_{\text{0}}$ denotes the intensity of the used total field scattered field (TFSF) source and P the measured power of the monitors that completely envelope the bowtie. Plotting $\upsigma$ as a function of the photon energy yields the spectral position of the electric dipole resonance energy, which is strongly influenced by the size\cite{muhlschlegel2005resonant,fischer2008engineering,hanke2009efficient,prangsma2012electrically}, feed gap\cite{fromm2004gap,schuck2005improving,merlein2008nanomechanical,hanke2009efficient}, shape\cite{sonnichsen2002drastic,fischer2008engineering}, and dielectric environment\cite{maier2007plasmonics} of the plasmonic dimer. Typical results for s\,=\,110\,nm, g\,=\,10\,nm and varying t are presented in fig. \ref{sample}(c). With decreasing t from an initial value of t\,=\,60\,nm down to t\,=\,35\,nm in steps of 5\,nm, we observe an expected decrease of $\upsigma$ from $\upsigma$\,=\,\SI{0.051}{\micro\meter^2} to $\upsigma$\,=\,\SI{0.040}{\micro\meter^2}. This value, however, is still 2.26$\times$ larger than the geometrical area of the nanoantenna as can be seen on the normalized scale on the right axis of fig. \ref{sample}(c). In addition, we obtained a redshift of the SPR peak position from 1.33\,eV down to 1.29\,eV when decreasing the metal film thickness. For even smaller t, the redshift becomes more prominent and leads to SPR peak at 0.80\,eV for 5\,nm thick structures. In order to achieve the best resolution during the electron beam lithography, resulting in sharp tips and small feed gaps and, therefore, a high electric field enhancement, the structures should be as thin as possible\cite{rai1997handbook}. Taking into account the measured redshift of 0.25\,$\pm$\,0.05\,eV introduced by the GaAs substrate (see below), an additional redshift of 0.3-0.5\,eV by a very thin structure would require very small particle sizes, of the order of 50\,nm, to match the SPR and the quantum dot's emission range. Smaller particles, however, show a lower scattering to absorption ratio\cite{maier2007plasmonics} and are, therefore, not ideal. Hence, we chose a Au thickness of 35\,nm representing the best trade-off between high resolution and optimum scattering properties.   

Fig. \ref{sample}(d) and (e) show a typical scanning electron microscopy (SEM) image of the fabricated arrays and a close-up of a single bowtie on GaAs, respectively. Using a 35\,nm thick Au-film we could reproducibly fabricate feed-gaps and tip radii (r) as small as 10\,nm with a yield of almost 100\,\% even without using an adhesion layer. All triangles are equilateral and we define the size of a bowtie as the height of one individual nanotriangle composing the bowtie antenna. To study the influence of the high refractive index substrate ($\text{n}_{\text{GaAs}}$\,=\,3.54 @ T\,=297\,K and $\text{E}_{\text{Photon}}$\,=\,1.3\,eV\cite{blakemore1982semiconducting}) in more detail, we also fabricated reference structures on glass substrates ($\text{n}_{\text{glass}}$\,=\,1.52 @ $\text{E}_{\text{Photon}}$\,=\,2.1\,eV\cite{2014}), as shown in the SEM image in fig. \ref{sample}(f). The geometrical properties are nominally identical to the ones on GaAs except for the presence of a 5\,nm thick Titanium adhesion layer below the 35\,nm Au-film. The achieved resolution is slightly reduced due to the non-conductive substrate which is disadvantageous for the electron beam lithography. Further details on the fabrication process are presented in the methods section.

To optically probe the SPR and scattering cross section of our structures we used a room temperature white light $\upmu$-reflectivity setup. Light from a halogen lamp was polarized along the bowtie axis $(\Theta = 0^\circ)$, focused on the sample surface, spectrally analyzed in a 0.5\,m spectrometer and detected with a liquid nitrogen cooled charge coupled device (CCD) camera. Thereby, we recorded a reflectivity spectrum from the bowtie array (S\textsuperscript{BT}($\upomega$)) and a reference spectrum from the bare substrate directly adjacent to the bowtie array (S\textsuperscript{ref}($\upomega$)). We then normalize the two data sets according to I($\upomega$)\,=\,[$\frac{\text{S\textsuperscript{BT}($\upomega$)}}{\text{S\textsuperscript{ref}($\upomega$)}}-1$] representing a measure of the reflectivity change caused by the bowties. This method reveals the scattering spectrum and, therefore, the SPR frequency of the probed structures. The spot size was determined to be 9\,$\upmu$m, such that we probe $\sim$\,30 bowties simultaneously leading to good statistics with a single measurement. However, as we observe small fabrication imperfection ($\leq$\,10\,nm) from SEM images, it is highly likely that all spectra are inhomogeneously broadened.


%
%

%
%
\section{Results and Discussion}

%
%
\begin{figure*}[htbp]
\includegraphics[width=\textwidth]{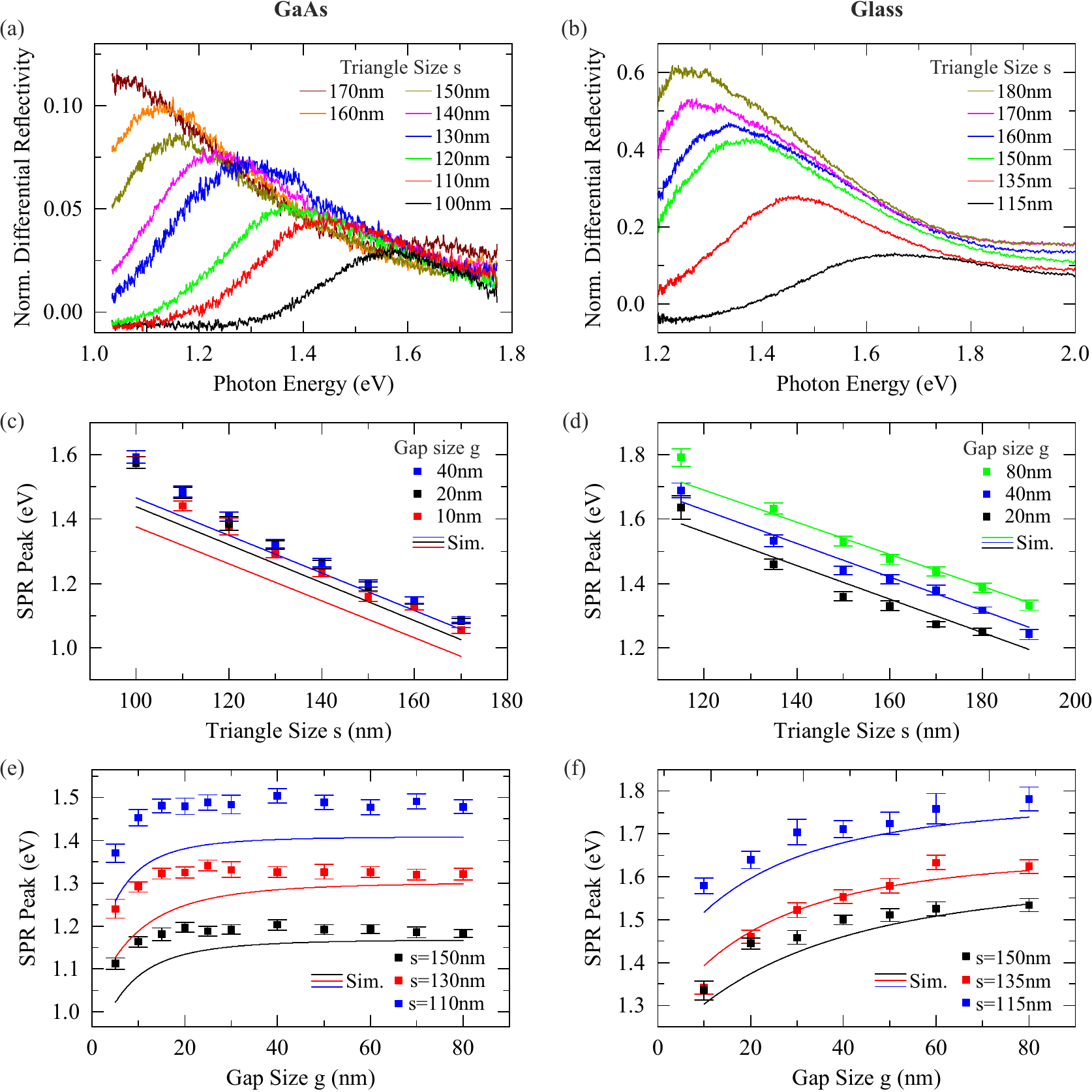}
\caption{\label{SPR} Comparison of surface plasmon resonance properties between GaAs (left row) and glass (right row).  (a),(b) Typical SPR spectra for bowties with g\,=\,20\,nm and different triangle sizes. (c),(d) SPR peak energy as a function of the triangle sizes for different feed gaps g: 10\,nm (red), 20\,nm (black), 40\,nm (blue) and 80\,nm (green). Solid lines are corresponding FDTD simulations. (e),(f) SPR peak energy as a function of feed gap size g for different triangle sizes s: 110\,nm and 115\,nm (blue), 130\,nm and 135\,nm (red) and 150\,nm (black).}
\end{figure*}

A detailed study of the plasmonic response of the nanoanteannas for different geometrical parameters is presented in this section. Although the focus is on the high refractive index, semiconducting GaAs substrate, we compare our results to these measured for similar structures on a glass substrate. Typical normalized differential reflectivity spectra obtained for bowtie arrays on GaAs with g\,=\,20\,$\pm$\,5\,nm and varying triangle sizes are shown in fig. \ref{SPR}(a). As expected, the scattering cross section and, therefore, the measured relative differential intensity decreases with decreasing structure sizes, whereas the SPR peak energy increases\cite{maier2007plasmonics}. Compared to the spectra recorded on glass, shown in fig. \ref{SPR}(b), the relative intensity is $\sim$5$\times$ lower due to the enhanced reflectivity of the GaAs substrate, e.g. 0.08 compared to 0.42 for s\,=\,150\,nm. For both substrates the resonances have a full width at half maximum (FWHM) of 0.4\,$\pm$\,0.1\,eV corresponding to a dephasing time of 3.3\,$\pm$\,0.8\,fs, in good agreement with values reported in the literature\cite{sonnichsen2002drastic,hanke2009efficient}. 

The SPR peak energy as a function of the triangle size is plotted in fig. \ref{SPR}(c) and (d) for GaAs and glass, respectively. For g\,=\,20\,nm, we observe a linear shift from 1.57\,$\pm$\,0.02\,eV to 1.08\,$\pm$\,0.01\,eV when changing the triangles size from 100\,nm to 170 \,nm on GaAs. These values translate to a shift rate of 7.0\,$\pm$\,0.5\,meV/nm. A qualitatively similar trend is observed on glass. Here, the SPR shifts from 1.63\,$\pm$\,0.04\,eV to 1.25\,$\pm$\,0.01\,eV when tuning the triangle size from 115\,nm to 180 \,nm, corresponding to a shift rate of 5.9\,$\pm$\,0.8\,meV/nm.  Thus, we found shift rates which are similar within the error and a red shift of the SPR by 0.25\,$\pm$\,0.05\,eV between the different substrates for s\,=\,150\,nm and g\,=\,20\,nm. This observation is attributed to the higher refractive index of GaAs\cite{maier2007plasmonics} as compared to glass. All results obtained on glass are supported quantitatively by our FDTD simulations, whereas on GaAs we find good qualitative agreement. For all simulations, we used a triangle tip radius r\,=\,20\,nm instead of the experimental observed 10\,nm. This is not expected to have any strong quantitative impact on our simulation results due to inhomogeneities of the triangle size ($\pm$\,5\,\%) that dominate the SPR frequency. 
We note that the native oxide layer on top of our GaAs wafers is included in the simulation since it strongly influences the plasmonic properties\cite{davies2012metal} due to its much lower refractive index $\text{n}_{\text{oxide}}$\,$\sim$\,1.5 as compared to the GaAs substrate $\text{n}_{\text{GaAs}}$\,$\sim$\,3.5. By selectively etching the oxide away at a certain region of the sample and performing atomic force microscopy measurements, we determined the thickness of the oxide layer to be 3.5\,$\pm$\,1\,nm, in very good agreement with values reported in the literature \cite{davies2012metal}. From our simulations (data not shown) we expect a blue shift of the SPR on GaAs by 0.18\,eV due to the presence of a 4\,nm thin oxide layer. 

Another possibility to influence the SPR is to vary the feed gap. This leads to a red shift of the SPR with decreasing gap for both substrates due to the increased coupling between the triangles. This mechanism lowers the effective restoring force of the oscillating free electron plasma in the nanoparticles and, therefore, decreases the resonance energy\cite{maier2007plasmonics,Jain2011}. To investigate this coupling effect in more detail, we experimentally and theoretically studied the SPR as a function of g for different triangle sizes. The results obtained on GaAs are plotted in fig. \ref{SPR}(e). All curves follow a g\textsuperscript{-3}-dependence, which can be derived from the simple qualitative picture of two interacting dipoles\cite{vollmer1995optical}. This behavior, which originates from the cubic decay of the near field of a point dipole\cite{Jackson1998}, is also measured on a glass substrate.  However, we observe a clear difference between the two material systems. Whilst for GaAs the SPR only starts to shift when the gap becomes smaller than g\,=\,20\,nm, we already observe a change at g\,=\,50\,nm for glass substrates. Furthermore, the absolute shift of 0.20\,eV when decreasing the gap from g\,=\,80\,nm to g\,=\,10\,nm is almost one order of magnitude lager for glass compared to 0.03\,eV for GaAs. All experimental observations are again confirmed by our FDTD simulations, which agree well with the measured data (solid lines - fig. \ref{SPR}(e) and (f)). The obtained results indicate a weaker coupling between the individual bowtie triangles on GaAs. This could be related to increased damping of the surface plasmon due to the higher refractive index substrate. However, we believe that this is not fully responsible for the reduction of the coupling strength by one order of magnitude since the SPR linewidth and, therefore, the plasmon lifetime found in fig. \ref{SPR}(a) is similar for both substrates.

\begin{figure*}[htbp]
\includegraphics[width=\textwidth]{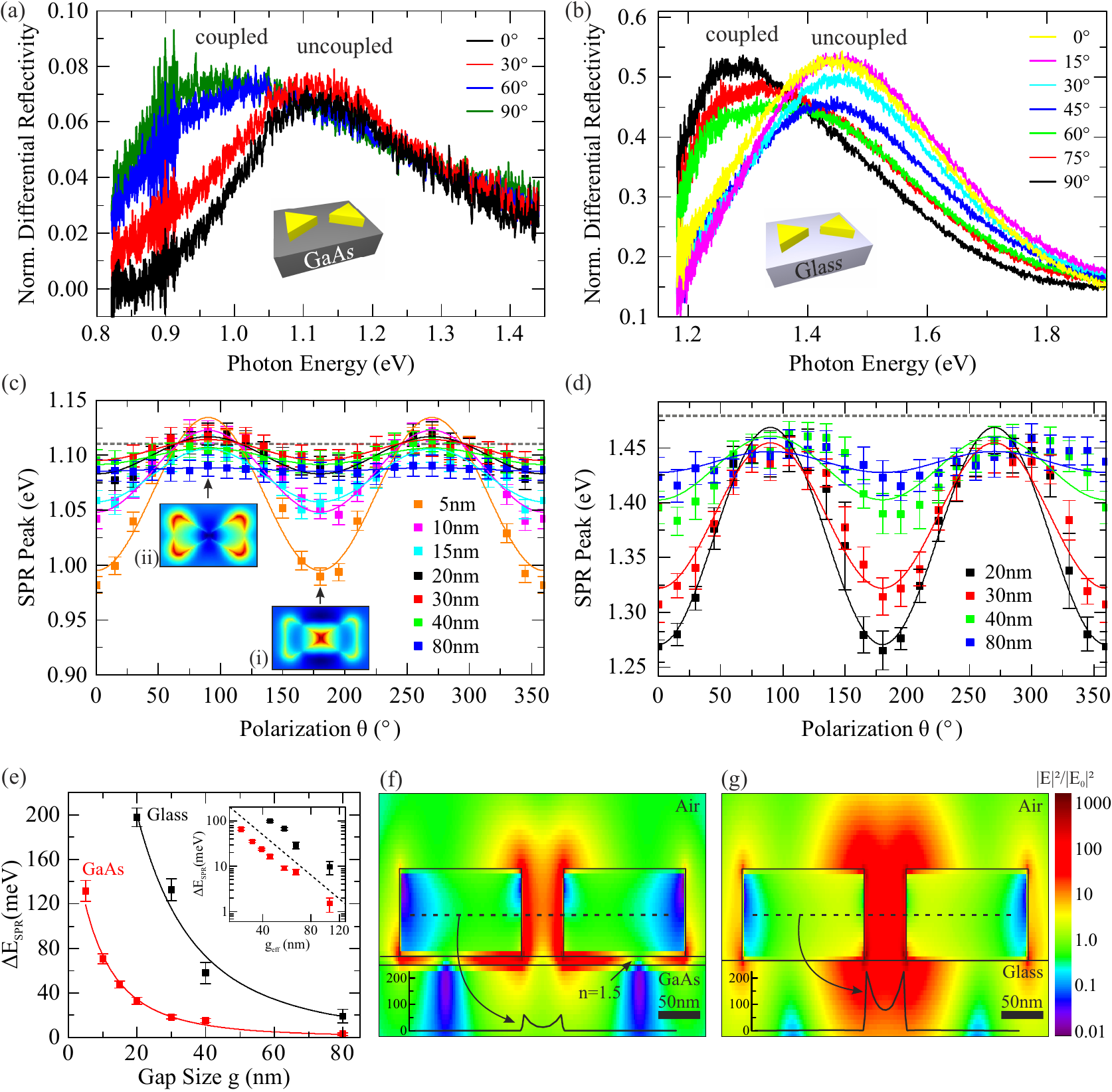}
\caption{\label{pol} Normalized differential reflectivity spectra for different polarization angles for (a) bowties on GaAs with s\,=\,170\,nm and g\,=\,5\,nm and (b) glass with s\,=\,170\,nm and g\,=\,20\,nm. (c),(d) Corresponding SPR peak positions as a function of polarization angle for bowties on GaAs (c) and glass (d) with various gap sizes. Solid lines are sin\textsuperscript{2}-fits. Insets show simulated electric field enhancement for (i) $\Theta$\,=\,0$^\circ$ and (ii) $\Theta$\,=\,90$^\circ$. (e) $\Updelta\text{E}_{\text{SPR}}$ as a function of gap size g. (f),(g) Simulated electric field enhancements for bowties (s\,=\,150\,nm, g\,=\,50\,nm) on GaAs and glass, respectively, on a logarithmic scale. Insets show cross section along the dashed line at y\,=\,0\,nm on a linear scale.}
\end{figure*}

To gain deeper insight into the interaction behavior of the individual triangles, we varied the polarization axis of the incident white light and explored the impact on the energetic position of the SPR. All measurements presented above were obtained with the polarization aligned along the long bowtie axis ($\Theta$\,=\,0$^\circ$), which means that the induced charge oscillation pushes the electrons towards the tips at the feed gap (see inset (i), fig. \ref{pol}(c)). If the gap is much smaller than the wavelength of the driving field, a significant fraction of the plasmon near-field can reach the adjacent triangle and lower the restoring force of the free electron plasma, resulting in a redshift of the SPR\cite{maier2007plasmonics}. In contrast, the electrons are pushed into the non-facing tips of the bowtie if the polarization is rotated by 90$^\circ$ (see inset (ii), fig. \ref{pol}(c)). Due to their relatively large separation, the near-fields cannot interact strongly and the SPR frequency is close to that of a single uncoupled triangle ($\text{E}^{\text{u.c.}}_{\text{SPR}}$). 

In fig. \ref{pol}(a) and (b), we present polarization resolved differential reflectivity spectra for GaAs (s\,=\,170\,nm, g \,=\,5\,nm) and glass (s\,=\,170\,nm, g \,=\,20\,nm) samples, respectively. The shift from the coupled to the uncoupled mode is clearly visible for both substrates. Furthermore, we observe a broadening of the SPR's FWHM by $\sim$\,50\,$\pm$\,10\,meV for the coupled case, which we attribute to increased radiation damping. This effect only depends on the volume of the probed structure\cite{wokaun1982radiation}, which is twice as large for the coupled mode than for the uncoupled one. In fig. \ref{pol}(c), the SPR peak energy is plotted as a function of the polarization angle $\Theta$ of the incident light for bowties on GaAs with s\,=\,170\,nm and g ranging from 5\,nm to 80\,nm. When the polarization is tuned perpendicular to the bowtie axis ($\Theta$\,=\,90$^\circ$ and $\Theta$\,=\,270$^\circ$), we observe values of 1.10\,$\pm$\,0.01\,eV, close to the resonance energy of uncoupled, nominal identical triangles. The grey dashed line indicates the position of the SPR peak energy for a single triangle, obtained from our FDTD simulations. Furthermore, the inset (ii) shows the corresponding electric field distribution where we observe the two independent modes of the individual triangles. In contrast, when we turn the polarization parallel to the long bowtie axis ($\Theta$\,=\,0$^\circ$ and $\Theta$\,=\,180$^\circ$), we probe the coupled mode (inset (i)) and obtain a red shift, the size of which peaks at $\Delta \text{E}_{\text{SPR}}$\,=\,0.14\,$\pm$\,0.01\,eV for g\,=\,5\,nm. This continuous change in peak energy can be well described by $\text{E}_\text{SPR}(\Theta)=\text{E}^{\text{u.c.}}_{\text{SPR}}-\Delta \text{E}_{\text{SPR}}\cdot \sin^2 (\Theta)$ (solid lines in fig. \ref{pol}(c) and (d)) whose amplitude $\Delta \text{E}_{\text{SPR}}$ increases with decreasing gap and increasing coupling, respectively. The same behavior is found for nominally identical bowties on a glass substrate, shown in fig. \ref{pol}(d). However, we detect a redshift of $\Delta \text{E}_{\text{SPR}}$\,=\,0.20\,$\pm$\,0.01\,eV already at g\,=\,20\,nm supporting our expectation of enhanced interaction between the individual bowtie triangles on glass as compared to GaAs. To quantify this behavior, we plotted the obtained $\Delta \text{E}_{\text{SPR}}$ as a function of g as shown in fig. \ref{pol}(e). As previously mentioned, on the glass substrate there is already a significant coupling effect for g\,=\,50\,nm, whereas on GaAs the triangles show a relevant interaction only for g\,$\leq$\,20\,nm. The two curves follow again a g\textsuperscript{-3}-trend, indicating that the coupling can be visualized as a dipole-dipole interaction\cite{Jain2011}. In this simplified picture we treat the triangles as two point dipoles which are separated by g plus an additional offset ${\text{g}}_{\text{0}}$ that depends on the charge distribution inside the triangles. The resulting fit formula reads then:
\begin{equation}
\Delta \text{E}_{\text{SPR}} = C\cdot(g+g_0)^{-3}
\label{formel}
\end{equation}

where C determines the curvature and, therefore, the coupling strength which strongly depends on the used geometry and substrate. From the fit of the measurement data, we obtained a value for $\text{g}_{\text{0,glass}}$\,=\,30\,$\pm$\,9\,nm and $\text{g}_{\text{0,GaAs}}$\,=\,24\,$\pm$\,2\,nm, identical within the experimental error. For C we obtained $\text{C}_{\text{glass}}$\,=\,12.5\,$\pm$\,6.6\,$\frac{keV}{nm^3}$ and $\text{C}_{\text{GaAs}}$\,=\,1.5\,$\pm$\,0.3\,$\frac{keV}{nm^3}$ indicating that the coupling between the triangles on glass substrates is $\sim$8\,$\times$ stronger than on GaAs for comparable geometric parameters. In the inset of fig. \ref{pol}(e) we plotted the same data as a function of the effective separation between the two dipoles $\text{g}_{\text{eff}}$\,=\,$\text{g}$\,+\,$\text{g}_{\text{0,mean}}$ on a double logarithmic scale. For the offset we used $\text{g}_{\text{0,mean}}$\,=\,24.5\,nm, the weighted mean value obtained from our fits. As a guide to the eye, we also plotted a dashed line having a slope of -3, indicating that we indeed observe a g\textsuperscript{-3}-trend in our measurements. The origin of the pronounced difference between the two material systems becomes clear upon looking at simulations of the electric field intensity around the bowtie (s\,=\,150\,nm, g\,=\,50\,nm) at the SPR frequency as shown in fig. \ref{pol}(f) and (g) on a logarithmic scale for GaAs and glass, respectively. In the case of glass, most of the electromagnetic energy is located in and around the feed gap of the antenna. Moreover, if g\,\textless\,80\,nm the fields of the individual triangles penetrate into the neighboring nanotriangle and interact with the free electron plasma. In contrast, the electric field intensity in the GaAs samples is more strongly localized directly at the gold surface in the feed gap and especially in the oxide layer between the gold and the GaAs. This leads to a decrease in the coupling strength between the two triangles compared to the identical structures on glass. The inset of both figures show a cross section along the dashed line at y\,=\,0\,nm on a linear scale. We found that the field exactly in the middle of the feed gap is 14\,$\times$ enhanced on GaAs compared to 79\,$\times$ for nominally identical triangles on glass. Also the exponential decay of the electric field intensity within the feed gap is faster on GaAs (5.2\,$\pm$\,0.3\,nm) than on glass (7.8\,$\pm$\,0.2\,nm). From those findings we conclude that the lower coupling in the GaAs samples can be explained by a lower overlap of the electric fields between the two triangles. 

It is remarkable that the calculated field enhancement in the GaAs samples is largest at the gold/oxide interface where enhancement factors up to 570\,$\times$ were found as compared to 90\,$\times$ at the gold/air interface. In future experiments, this strong field enhancement could be used in optically active plasmonic-semiconducting systems, where bowtie antennas are coupled to proximal active emitters such as InGaAs quantum dots in order to tailor their emissive properties\cite{doi:10.1021/nl102548t,bracher2014optical}. Furthermore, we point out that by addressing single bowties, further decreasing the gap size, and using monocrystalline gold\cite{huang2010atomically,prangsma2012electrically} it should be possible, especially on a glass substrate, to reach a regime where the splitting between the coupled and uncoupled mode of a bowtie is bigger than their linewidth. This tunable and significant coupling between the two orthogonally polarized plasmonic modes may open the way toward THz spectroscopy and parametric coherent driving of isolated nano objects placed into the feed gap\cite{giannini2010scattering,raithel1997compression}. Moreover, the use of a semiconductor substrate as demonstrated in this study may even facilitate such experiments on individual quantum emitters that have already demonstrated excellent coherence properties. It could, therefore, be possible to link the THz and optical regimes coherently at the quantum limit.



\section{Conclusions}

In summary, we presented a comprehensive study of the optical properties of gold bowtie nanoantennas defined by electron beam lithography on GaAs substrates. Using FDTD simulations, we determined the optimum Au thickness of our structures to be $\sim$\,35\,nm representing a tradeoff between good scattering properties and structures having small feed gaps and sharp tip radii. We fabricated bowtie nanoantennas with sizes s\,=\,100-190\,nm, feed gaps g\,=\,5-80\,nm and tip radii of the order of 10\,nm on GaAs and glass. The SPR peak energy for bowtie antennas on GaAs samples was found to red shift linearly with increasing size at a rate of 7.0\,$\pm$\,0.5\,meV/nm and can, therefore, be tuned through the emission range of self-assembled InGaAs quantum dots around 1.3\,eV. We found a uniform global redshift of the SPR of 0.25\,$\pm$\,0.05\,eV on GaAs compared to the samples on glass. Gap dependent measurements showed a clear difference in the coupling strength, as we observed a redshift of 0.03\,eV on GaAs when decreasing the feed gap from 80\,nm to 10\,nm compared to 0.20\,eV for the glass sample. Using polarization resolved measurements, we quantified the coupling strength to be $\sim$8 times lower on GaAs as compared to glass. From our simulations, which support our obtained results, we conclude that this effect is caused by a modification of the electric field distribution due to the difference of the substrate's refractive indices and the presence of a 4\,nm thin native oxide layer on top of the GaAs wafer. The obtained results provide important information for the integration of plasmonic nanoantennas in novel, photonic, on-chip devices and the design of future plasmonic hybrid systems.

\begin{acknowledgments}
We acknowledge financial support of the DFG via the SFB 631, Teilprojekt B3, the German Excellence Initiative via NIM, and FP-7 of the European Union via SOLID. The author gratefully acknowledges the support of the TUM International Graduate School of Science and Engineering (IGSSE).
\end{acknowledgments}

\begin{appendix}
\section{Methods}

The samples investigated were defined on undoped GaAs [100] wafers or glass (MENZEL microscope cover slips) substrates. After cleavage, the samples were flushed with acetone and isopropanol (IPA). In order to get a better adhesion of the e-beam resist, the samples were put on a hot plate (170\,$^\circ$C) for 5\,min. An e-beam resist (Polymethylmethacrylat 950K, AR-P 679.02, ALLRESIST) was coated at 4000\,rpm  for 40\,s at an acceleration of 2000\,rpm/s and baked out at 170~$^\circ$C for 5\,min, producing a resist thickness of $70\pm5$\,nm. For the glass samples, we evaporated 10\,nm aluminum on top of the PMMA layer to avoid charging effects during the e-beam writing. The samples were illuminated in a Raith E-line system using an acceleration voltage of 30\,kV and an aperture of 10\,$\mu$m. A dose test was performed for every fabrication run, as this crucial parameter depends on the varying electron beam current. Typical values were 800\,$\mu$As/cm\textsuperscript{2} for GaAs and 700\,$\mu$As/cm\textsuperscript{2} for glass substrates. After the electron beam writing the Al layer on the glass samples was etched away using a metal-ion-free photoresist developer (AZ 726 MIF, MicroChemicals). All samples were developed in Methylisobutylketon (MIBK) diluted with IPA (1:3) for 45\,s. To stop the development, the sample was rinsed with pure IPA. For the metalization an electron beam evaporator was used to deposit a 5\,nm Titanium adhesion layer for the glass and 35\,nm gold for all substrates at a low rate of 1\,\AA/s. The lift-off was performed in 50\,$^\circ$C warm acetone, leaving behind high quality nanostructures with features sizes in the order of 10\,nm. 

\end{appendix}


\begin{thebibliography}{49}%
\makeatletter
\providecommand \@ifxundefined [1]{%
 \@ifx{#1\undefined}
}%
\providecommand \@ifnum [1]{%
 \ifnum #1\expandafter \@firstoftwo
 \else \expandafter \@secondoftwo
 \fi
}%
\providecommand \@ifx [1]{%
 \ifx #1\expandafter \@firstoftwo
 \else \expandafter \@secondoftwo
 \fi
}%
\providecommand \natexlab [1]{#1}%
\providecommand \enquote  [1]{``#1''}%
\providecommand \bibnamefont  [1]{#1}%
\providecommand \bibfnamefont [1]{#1}%
\providecommand \citenamefont [1]{#1}%
\providecommand \href@noop [0]{\@secondoftwo}%
\providecommand \href [0]{\begingroup \@sanitize@url \@href}%
\providecommand \@href[1]{\@@startlink{#1}\@@href}%
\providecommand \@@href[1]{\endgroup#1\@@endlink}%
\providecommand \@sanitize@url [0]{\catcode `\\12\catcode `\$12\catcode
  `\&12\catcode `\#12\catcode `\^12\catcode `\_12\catcode `\%12\relax}%
\providecommand \@@startlink[1]{}%
\providecommand \@@endlink[0]{}%
\providecommand \url  [0]{\begingroup\@sanitize@url \@url }%
\providecommand \@url [1]{\endgroup\@href {#1}{\urlprefix }}%
\providecommand \urlprefix  [0]{URL }%
\providecommand \Eprint [0]{\href }%
\providecommand \doibase [0]{http://dx.doi.org/}%
\providecommand \selectlanguage [0]{\@gobble}%
\providecommand \bibinfo  [0]{\@secondoftwo}%
\providecommand \bibfield  [0]{\@secondoftwo}%
\providecommand \translation [1]{[#1]}%
\providecommand \BibitemOpen [0]{}%
\providecommand \bibitemStop [0]{}%
\providecommand \bibitemNoStop [0]{.\EOS\space}%
\providecommand \EOS [0]{\spacefactor3000\relax}%
\providecommand \BibitemShut  [1]{\csname bibitem#1\endcsname}%
\let\auto@bib@innerbib\@empty
\bibitem [{\citenamefont {Barnes}\ \emph {et~al.}(2003)\citenamefont {Barnes},
  \citenamefont {Dereux},\ and\ \citenamefont {Ebbesen}}]{barnes2003surface}%
  \BibitemOpen
  \bibfield  {author} {\bibinfo {author} {\bibfnamefont {W.~L.}\ \bibnamefont
  {Barnes}}, \bibinfo {author} {\bibfnamefont {A.}~\bibnamefont {Dereux}}, \
  and\ \bibinfo {author} {\bibfnamefont {T.~W.}\ \bibnamefont {Ebbesen}},\
  }\href@noop {} {\bibfield  {journal} {\bibinfo  {journal} {Nature}\ }\textbf
  {\bibinfo {volume} {424}},\ \bibinfo {pages} {824} (\bibinfo {year}
  {2003})}\BibitemShut {NoStop}%
\bibitem [{\citenamefont {Schuller}\ \emph {et~al.}(2010)\citenamefont
  {Schuller}, \citenamefont {Barnard}, \citenamefont {Cai}, \citenamefont
  {Jun}, \citenamefont {White},\ and\ \citenamefont
  {Brongersma}}]{schuller2010plasmonics}%
  \BibitemOpen
  \bibfield  {author} {\bibinfo {author} {\bibfnamefont {J.~A.}\ \bibnamefont
  {Schuller}}, \bibinfo {author} {\bibfnamefont {E.~S.}\ \bibnamefont
  {Barnard}}, \bibinfo {author} {\bibfnamefont {W.}~\bibnamefont {Cai}},
  \bibinfo {author} {\bibfnamefont {Y.~C.}\ \bibnamefont {Jun}}, \bibinfo
  {author} {\bibfnamefont {J.~S.}\ \bibnamefont {White}}, \ and\ \bibinfo
  {author} {\bibfnamefont {M.~L.}\ \bibnamefont {Brongersma}},\ }\href@noop {}
  {\bibfield  {journal} {\bibinfo  {journal} {Nature Materials}\ }\textbf
  {\bibinfo {volume} {9}},\ \bibinfo {pages} {193} (\bibinfo {year}
  {2010})}\BibitemShut {NoStop}%
\bibitem [{\citenamefont {Novotny}\ and\ \citenamefont {van
  Hulst}(2011)}]{novotny2011antennas}%
  \BibitemOpen
  \bibfield  {author} {\bibinfo {author} {\bibfnamefont {L.}~\bibnamefont
  {Novotny}}\ and\ \bibinfo {author} {\bibfnamefont {N.}~\bibnamefont {van
  Hulst}},\ }\href@noop {} {\bibfield  {journal} {\bibinfo  {journal} {Nature
  Photonics}\ }\textbf {\bibinfo {volume} {5}},\ \bibinfo {pages} {83}
  (\bibinfo {year} {2011})}\BibitemShut {NoStop}%
\bibitem [{\citenamefont {Gramotnev}\ and\ \citenamefont
  {Bozhevolnyi}(2014)}]{gramotnev2014nanofocusing}%
  \BibitemOpen
  \bibfield  {author} {\bibinfo {author} {\bibfnamefont {D.~K.}\ \bibnamefont
  {Gramotnev}}\ and\ \bibinfo {author} {\bibfnamefont {S.~I.}\ \bibnamefont
  {Bozhevolnyi}},\ }\href@noop {} {\bibfield  {journal} {\bibinfo  {journal}
  {Nature Photonics}\ }\textbf {\bibinfo {volume} {8}},\ \bibinfo {pages} {13}
  (\bibinfo {year} {2014})}\BibitemShut {NoStop}%
\bibitem [{\citenamefont {Kim}\ \emph {et~al.}(2008)\citenamefont {Kim},
  \citenamefont {Jin}, \citenamefont {Kim}, \citenamefont {Park}, \citenamefont
  {Kim},\ and\ \citenamefont {Kim}}]{kim2008high}%
  \BibitemOpen
  \bibfield  {author} {\bibinfo {author} {\bibfnamefont {S.}~\bibnamefont
  {Kim}}, \bibinfo {author} {\bibfnamefont {J.}~\bibnamefont {Jin}}, \bibinfo
  {author} {\bibfnamefont {Y.}~\bibnamefont {Kim}}, \bibinfo {author}
  {\bibfnamefont {I.}~\bibnamefont {Park}}, \bibinfo {author} {\bibfnamefont
  {Y.}~\bibnamefont {Kim}}, \ and\ \bibinfo {author} {\bibfnamefont
  {S.}~\bibnamefont {Kim}},\ }\href@noop {} {\bibfield  {journal} {\bibinfo
  {journal} {Nature}\ }\textbf {\bibinfo {volume} {453}},\ \bibinfo {pages}
  {757} (\bibinfo {year} {2008})}\BibitemShut {NoStop}%
\bibitem [{\citenamefont {Hanke}\ \emph {et~al.}(2012)\citenamefont {Hanke},
  \citenamefont {Cesar}, \citenamefont {Knittel}, \citenamefont {Tr{\"u}gler},
  \citenamefont {Hohenester}, \citenamefont {Leitenstorfer},\ and\
  \citenamefont {Bratschitsch}}]{hanke2012tailoring}%
  \BibitemOpen
  \bibfield  {author} {\bibinfo {author} {\bibfnamefont {T.}~\bibnamefont
  {Hanke}}, \bibinfo {author} {\bibfnamefont {J.}~\bibnamefont {Cesar}},
  \bibinfo {author} {\bibfnamefont {V.}~\bibnamefont {Knittel}}, \bibinfo
  {author} {\bibfnamefont {A.}~\bibnamefont {Tr{\"u}gler}}, \bibinfo {author}
  {\bibfnamefont {U.}~\bibnamefont {Hohenester}}, \bibinfo {author}
  {\bibfnamefont {A.}~\bibnamefont {Leitenstorfer}}, \ and\ \bibinfo {author}
  {\bibfnamefont {R.}~\bibnamefont {Bratschitsch}},\ }\href@noop {} {\bibfield
  {journal} {\bibinfo  {journal} {Nano Letters}\ }\textbf {\bibinfo {volume}
  {12}},\ \bibinfo {pages} {992} (\bibinfo {year} {2012})}\BibitemShut
  {NoStop}%
\bibitem [{\citenamefont {Punj}\ \emph {et~al.}(2013)\citenamefont {Punj},
  \citenamefont {Mivelle}, \citenamefont {Moparthi}, \citenamefont {van
  Zanten}, \citenamefont {Rigneault}, \citenamefont {van Hulst}, \citenamefont
  {Garc{\'\i}a-Paraj{\'o}},\ and\ \citenamefont {Wenger}}]{punj2013plasmonic}%
  \BibitemOpen
  \bibfield  {author} {\bibinfo {author} {\bibfnamefont {D.}~\bibnamefont
  {Punj}}, \bibinfo {author} {\bibfnamefont {M.}~\bibnamefont {Mivelle}},
  \bibinfo {author} {\bibfnamefont {S.~B.}\ \bibnamefont {Moparthi}}, \bibinfo
  {author} {\bibfnamefont {T.~S.}\ \bibnamefont {van Zanten}}, \bibinfo
  {author} {\bibfnamefont {H.}~\bibnamefont {Rigneault}}, \bibinfo {author}
  {\bibfnamefont {N.~F.}\ \bibnamefont {van Hulst}}, \bibinfo {author}
  {\bibfnamefont {M.~F.}\ \bibnamefont {Garc{\'\i}a-Paraj{\'o}}}, \ and\
  \bibinfo {author} {\bibfnamefont {J.}~\bibnamefont {Wenger}},\ }\href@noop {}
  {\bibfield  {journal} {\bibinfo  {journal} {Nature Nanotechnology}\ }\textbf
  {\bibinfo {volume} {8}},\ \bibinfo {pages} {512} (\bibinfo {year}
  {2013})}\BibitemShut {NoStop}%
\bibitem [{\citenamefont {Anger}\ \emph {et~al.}(2006)\citenamefont {Anger},
  \citenamefont {Bharadwaj},\ and\ \citenamefont
  {Novotny}}]{anger2006enhancement}%
  \BibitemOpen
  \bibfield  {author} {\bibinfo {author} {\bibfnamefont {P.}~\bibnamefont
  {Anger}}, \bibinfo {author} {\bibfnamefont {P.}~\bibnamefont {Bharadwaj}}, \
  and\ \bibinfo {author} {\bibfnamefont {L.}~\bibnamefont {Novotny}},\
  }\href@noop {} {\bibfield  {journal} {\bibinfo  {journal} {Physical Review
  Letters}\ }\textbf {\bibinfo {volume} {96}},\ \bibinfo {pages} {113002}
  (\bibinfo {year} {2006})}\BibitemShut {NoStop}%
\bibitem [{\citenamefont {K{\"u}hn}\ \emph {et~al.}(2006)\citenamefont
  {K{\"u}hn}, \citenamefont {H{\aa}kanson}, \citenamefont {Rogobete},\ and\
  \citenamefont {Sandoghdar}}]{kuhn2006enhancement}%
  \BibitemOpen
  \bibfield  {author} {\bibinfo {author} {\bibfnamefont {S.}~\bibnamefont
  {K{\"u}hn}}, \bibinfo {author} {\bibfnamefont {U.}~\bibnamefont
  {H{\aa}kanson}}, \bibinfo {author} {\bibfnamefont {L.}~\bibnamefont
  {Rogobete}}, \ and\ \bibinfo {author} {\bibfnamefont {V.}~\bibnamefont
  {Sandoghdar}},\ }\href@noop {} {\bibfield  {journal} {\bibinfo  {journal}
  {Physical Review Letters}\ }\textbf {\bibinfo {volume} {97}},\ \bibinfo
  {pages} {017402} (\bibinfo {year} {2006})}\BibitemShut {NoStop}%
\bibitem [{\citenamefont {Pompa}\ \emph {et~al.}(2006)\citenamefont {Pompa},
  \citenamefont {Martiradonna}, \citenamefont {Della~Torre}, \citenamefont
  {Della~Sala}, \citenamefont {Manna}, \citenamefont {De~Vittorio},
  \citenamefont {Calabi}, \citenamefont {Cingolani},\ and\ \citenamefont
  {Rinaldi}}]{pompa2006metal}%
  \BibitemOpen
  \bibfield  {author} {\bibinfo {author} {\bibfnamefont {P.}~\bibnamefont
  {Pompa}}, \bibinfo {author} {\bibfnamefont {L.}~\bibnamefont {Martiradonna}},
  \bibinfo {author} {\bibfnamefont {A.}~\bibnamefont {Della~Torre}}, \bibinfo
  {author} {\bibfnamefont {F.}~\bibnamefont {Della~Sala}}, \bibinfo {author}
  {\bibfnamefont {L.}~\bibnamefont {Manna}}, \bibinfo {author} {\bibfnamefont
  {M.}~\bibnamefont {De~Vittorio}}, \bibinfo {author} {\bibfnamefont
  {F.}~\bibnamefont {Calabi}}, \bibinfo {author} {\bibfnamefont
  {R.}~\bibnamefont {Cingolani}}, \ and\ \bibinfo {author} {\bibfnamefont
  {R.}~\bibnamefont {Rinaldi}},\ }\href@noop {} {\bibfield  {journal} {\bibinfo
   {journal} {Nature Nanotechnology}\ }\textbf {\bibinfo {volume} {1}},\
  \bibinfo {pages} {126} (\bibinfo {year} {2006})}\BibitemShut {NoStop}%
\bibitem [{\citenamefont {Kinkhabwala}\ \emph {et~al.}(2009)\citenamefont
  {Kinkhabwala}, \citenamefont {Yu}, \citenamefont {Fan}, \citenamefont
  {Avlasevich}, \citenamefont {M{\"u}llen},\ and\ \citenamefont
  {Moerner}}]{kinkhabwala2009large}%
  \BibitemOpen
  \bibfield  {author} {\bibinfo {author} {\bibfnamefont {A.}~\bibnamefont
  {Kinkhabwala}}, \bibinfo {author} {\bibfnamefont {Z.}~\bibnamefont {Yu}},
  \bibinfo {author} {\bibfnamefont {S.}~\bibnamefont {Fan}}, \bibinfo {author}
  {\bibfnamefont {Y.}~\bibnamefont {Avlasevich}}, \bibinfo {author}
  {\bibfnamefont {K.}~\bibnamefont {M{\"u}llen}}, \ and\ \bibinfo {author}
  {\bibfnamefont {W.}~\bibnamefont {Moerner}},\ }\href@noop {} {\bibfield
  {journal} {\bibinfo  {journal} {Nature Photonics}\ }\textbf {\bibinfo
  {volume} {3}},\ \bibinfo {pages} {654} (\bibinfo {year} {2009})}\BibitemShut
  {NoStop}%
\bibitem [{\citenamefont {Acuna}\ \emph {et~al.}(2012)\citenamefont {Acuna},
  \citenamefont {M{\"o}ller}, \citenamefont {Holzmeister}, \citenamefont
  {Beater}, \citenamefont {Lalkens},\ and\ \citenamefont
  {Tinnefeld}}]{acuna2012fluorescence}%
  \BibitemOpen
  \bibfield  {author} {\bibinfo {author} {\bibfnamefont {G.}~\bibnamefont
  {Acuna}}, \bibinfo {author} {\bibfnamefont {F.}~\bibnamefont {M{\"o}ller}},
  \bibinfo {author} {\bibfnamefont {P.}~\bibnamefont {Holzmeister}}, \bibinfo
  {author} {\bibfnamefont {S.}~\bibnamefont {Beater}}, \bibinfo {author}
  {\bibfnamefont {B.}~\bibnamefont {Lalkens}}, \ and\ \bibinfo {author}
  {\bibfnamefont {P.}~\bibnamefont {Tinnefeld}},\ }\href@noop {} {\bibfield
  {journal} {\bibinfo  {journal} {Science}\ }\textbf {\bibinfo {volume}
  {338}},\ \bibinfo {pages} {506} (\bibinfo {year} {2012})}\BibitemShut
  {NoStop}%
\bibitem [{\citenamefont {Taminiau}\ \emph {et~al.}(2008)\citenamefont
  {Taminiau}, \citenamefont {Stefani}, \citenamefont {Segerink},\ and\
  \citenamefont {Van~Hulst}}]{taminiau2008optical}%
  \BibitemOpen
  \bibfield  {author} {\bibinfo {author} {\bibfnamefont {T.}~\bibnamefont
  {Taminiau}}, \bibinfo {author} {\bibfnamefont {F.}~\bibnamefont {Stefani}},
  \bibinfo {author} {\bibfnamefont {F.}~\bibnamefont {Segerink}}, \ and\
  \bibinfo {author} {\bibfnamefont {N.}~\bibnamefont {Van~Hulst}},\ }\href@noop
  {} {\bibfield  {journal} {\bibinfo  {journal} {Nature Photonics}\ }\textbf
  {\bibinfo {volume} {2}},\ \bibinfo {pages} {234} (\bibinfo {year}
  {2008})}\BibitemShut {NoStop}%
\bibitem [{\citenamefont {Curto}\ \emph {et~al.}(2010)\citenamefont {Curto},
  \citenamefont {Volpe}, \citenamefont {Taminiau}, \citenamefont {Kreuzer},
  \citenamefont {Quidant},\ and\ \citenamefont {van Hulst}}]{Curto20082010}%
  \BibitemOpen
  \bibfield  {author} {\bibinfo {author} {\bibfnamefont {A.~G.}\ \bibnamefont
  {Curto}}, \bibinfo {author} {\bibfnamefont {G.}~\bibnamefont {Volpe}},
  \bibinfo {author} {\bibfnamefont {T.~H.}\ \bibnamefont {Taminiau}}, \bibinfo
  {author} {\bibfnamefont {M.~P.}\ \bibnamefont {Kreuzer}}, \bibinfo {author}
  {\bibfnamefont {R.}~\bibnamefont {Quidant}}, \ and\ \bibinfo {author}
  {\bibfnamefont {N.~F.}\ \bibnamefont {van Hulst}},\ }\href {\doibase
  10.1126/science.1191922} {\bibfield  {journal} {\bibinfo  {journal}
  {Science}\ }\textbf {\bibinfo {volume} {329}},\ \bibinfo {pages} {930}
  (\bibinfo {year} {2010})}\BibitemShut {NoStop}%
\bibitem [{\citenamefont {Curto}\ \emph {et~al.}(2013)\citenamefont {Curto},
  \citenamefont {Taminiau}, \citenamefont {Volpe}, \citenamefont {Kreuzer},
  \citenamefont {Quidant},\ and\ \citenamefont {van
  Hulst}}]{curto2013multipolar}%
  \BibitemOpen
  \bibfield  {author} {\bibinfo {author} {\bibfnamefont {A.~G.}\ \bibnamefont
  {Curto}}, \bibinfo {author} {\bibfnamefont {T.~H.}\ \bibnamefont {Taminiau}},
  \bibinfo {author} {\bibfnamefont {G.}~\bibnamefont {Volpe}}, \bibinfo
  {author} {\bibfnamefont {M.~P.}\ \bibnamefont {Kreuzer}}, \bibinfo {author}
  {\bibfnamefont {R.}~\bibnamefont {Quidant}}, \ and\ \bibinfo {author}
  {\bibfnamefont {N.~F.}\ \bibnamefont {van Hulst}},\ }\href@noop {} {\bibfield
   {journal} {\bibinfo  {journal} {Nature Communications}\ }\textbf {\bibinfo
  {volume} {4}},\ \bibinfo {pages} {1750} (\bibinfo {year} {2013})}\BibitemShut
  {NoStop}%
\bibitem [{\citenamefont {M{\"u}hlschlegel}\ \emph {et~al.}(2005)\citenamefont
  {M{\"u}hlschlegel}, \citenamefont {Eisler}, \citenamefont {Martin},
  \citenamefont {Hecht},\ and\ \citenamefont
  {Pohl}}]{muhlschlegel2005resonant}%
  \BibitemOpen
  \bibfield  {author} {\bibinfo {author} {\bibfnamefont {P.}~\bibnamefont
  {M{\"u}hlschlegel}}, \bibinfo {author} {\bibfnamefont {H.}~\bibnamefont
  {Eisler}}, \bibinfo {author} {\bibfnamefont {O.}~\bibnamefont {Martin}},
  \bibinfo {author} {\bibfnamefont {B.}~\bibnamefont {Hecht}}, \ and\ \bibinfo
  {author} {\bibfnamefont {D.}~\bibnamefont {Pohl}},\ }\href@noop {} {\bibfield
   {journal} {\bibinfo  {journal} {Science}\ }\textbf {\bibinfo {volume}
  {308}},\ \bibinfo {pages} {1607} (\bibinfo {year} {2005})}\BibitemShut
  {NoStop}%
\bibitem [{\citenamefont {Fischer}\ and\ \citenamefont
  {Martin}(2008)}]{fischer2008engineering}%
  \BibitemOpen
  \bibfield  {author} {\bibinfo {author} {\bibfnamefont {H.}~\bibnamefont
  {Fischer}}\ and\ \bibinfo {author} {\bibfnamefont {O.~J.}\ \bibnamefont
  {Martin}},\ }\href@noop {} {\bibfield  {journal} {\bibinfo  {journal} {Optics
  Express}\ }\textbf {\bibinfo {volume} {16}},\ \bibinfo {pages} {9144}
  (\bibinfo {year} {2008})}\BibitemShut {NoStop}%
\bibitem [{\citenamefont {Hanke}\ \emph {et~al.}(2009)\citenamefont {Hanke},
  \citenamefont {Krauss}, \citenamefont {Tr{\"a}utlein}, \citenamefont {Wild},
  \citenamefont {Bratschitsch},\ and\ \citenamefont
  {Leitenstorfer}}]{hanke2009efficient}%
  \BibitemOpen
  \bibfield  {author} {\bibinfo {author} {\bibfnamefont {T.}~\bibnamefont
  {Hanke}}, \bibinfo {author} {\bibfnamefont {G.}~\bibnamefont {Krauss}},
  \bibinfo {author} {\bibfnamefont {D.}~\bibnamefont {Tr{\"a}utlein}}, \bibinfo
  {author} {\bibfnamefont {B.}~\bibnamefont {Wild}}, \bibinfo {author}
  {\bibfnamefont {R.}~\bibnamefont {Bratschitsch}}, \ and\ \bibinfo {author}
  {\bibfnamefont {A.}~\bibnamefont {Leitenstorfer}},\ }\href@noop {} {\bibfield
   {journal} {\bibinfo  {journal} {Physical Review Letters}\ }\textbf {\bibinfo
  {volume} {103}},\ \bibinfo {pages} {257404} (\bibinfo {year}
  {2009})}\BibitemShut {NoStop}%
\bibitem [{\citenamefont {Prangsma}\ \emph {et~al.}(2012)\citenamefont
  {Prangsma}, \citenamefont {Kern}, \citenamefont {Knapp}, \citenamefont
  {Grossmann}, \citenamefont {Emmerling}, \citenamefont {Kamp},\ and\
  \citenamefont {Hecht}}]{prangsma2012electrically}%
  \BibitemOpen
  \bibfield  {author} {\bibinfo {author} {\bibfnamefont {J.~C.}\ \bibnamefont
  {Prangsma}}, \bibinfo {author} {\bibfnamefont {J.}~\bibnamefont {Kern}},
  \bibinfo {author} {\bibfnamefont {A.~G.}\ \bibnamefont {Knapp}}, \bibinfo
  {author} {\bibfnamefont {S.}~\bibnamefont {Grossmann}}, \bibinfo {author}
  {\bibfnamefont {M.}~\bibnamefont {Emmerling}}, \bibinfo {author}
  {\bibfnamefont {M.}~\bibnamefont {Kamp}}, \ and\ \bibinfo {author}
  {\bibfnamefont {B.}~\bibnamefont {Hecht}},\ }\href@noop {} {\bibfield
  {journal} {\bibinfo  {journal} {Nano Letters}\ }\textbf {\bibinfo {volume}
  {12}},\ \bibinfo {pages} {3915} (\bibinfo {year} {2012})}\BibitemShut
  {NoStop}%
\bibitem [{\citenamefont {Fromm}\ \emph {et~al.}(2004)\citenamefont {Fromm},
  \citenamefont {Sundaramurthy}, \citenamefont {Schuck}, \citenamefont {Kino},\
  and\ \citenamefont {Moerner}}]{fromm2004gap}%
  \BibitemOpen
  \bibfield  {author} {\bibinfo {author} {\bibfnamefont {D.}~\bibnamefont
  {Fromm}}, \bibinfo {author} {\bibfnamefont {A.}~\bibnamefont
  {Sundaramurthy}}, \bibinfo {author} {\bibfnamefont {P.}~\bibnamefont
  {Schuck}}, \bibinfo {author} {\bibfnamefont {G.}~\bibnamefont {Kino}}, \ and\
  \bibinfo {author} {\bibfnamefont {W.}~\bibnamefont {Moerner}},\ }\href@noop
  {} {\bibfield  {journal} {\bibinfo  {journal} {Nano Letters}\ }\textbf
  {\bibinfo {volume} {4}},\ \bibinfo {pages} {957} (\bibinfo {year}
  {2004})}\BibitemShut {NoStop}%
\bibitem [{\citenamefont {Schuck}\ \emph {et~al.}(2005)\citenamefont {Schuck},
  \citenamefont {Fromm}, \citenamefont {Sundaramurthy}, \citenamefont {Kino},\
  and\ \citenamefont {Moerner}}]{schuck2005improving}%
  \BibitemOpen
  \bibfield  {author} {\bibinfo {author} {\bibfnamefont {P.}~\bibnamefont
  {Schuck}}, \bibinfo {author} {\bibfnamefont {D.}~\bibnamefont {Fromm}},
  \bibinfo {author} {\bibfnamefont {A.}~\bibnamefont {Sundaramurthy}}, \bibinfo
  {author} {\bibfnamefont {G.}~\bibnamefont {Kino}}, \ and\ \bibinfo {author}
  {\bibfnamefont {W.}~\bibnamefont {Moerner}},\ }\href@noop {} {\bibfield
  {journal} {\bibinfo  {journal} {Physical Review Letters}\ }\textbf {\bibinfo
  {volume} {94}},\ \bibinfo {pages} {17402} (\bibinfo {year}
  {2005})}\BibitemShut {NoStop}%
\bibitem [{\citenamefont {Merlein}\ \emph {et~al.}(2008)\citenamefont
  {Merlein}, \citenamefont {Kahl}, \citenamefont {Zuschlag}, \citenamefont
  {Sell}, \citenamefont {Halm}, \citenamefont {Boneberg}, \citenamefont
  {Leiderer}, \citenamefont {Leitenstorfer},\ and\ \citenamefont
  {Bratschitsch}}]{merlein2008nanomechanical}%
  \BibitemOpen
  \bibfield  {author} {\bibinfo {author} {\bibfnamefont {J.}~\bibnamefont
  {Merlein}}, \bibinfo {author} {\bibfnamefont {M.}~\bibnamefont {Kahl}},
  \bibinfo {author} {\bibfnamefont {A.}~\bibnamefont {Zuschlag}}, \bibinfo
  {author} {\bibfnamefont {A.}~\bibnamefont {Sell}}, \bibinfo {author}
  {\bibfnamefont {A.}~\bibnamefont {Halm}}, \bibinfo {author} {\bibfnamefont
  {J.}~\bibnamefont {Boneberg}}, \bibinfo {author} {\bibfnamefont
  {P.}~\bibnamefont {Leiderer}}, \bibinfo {author} {\bibfnamefont
  {A.}~\bibnamefont {Leitenstorfer}}, \ and\ \bibinfo {author} {\bibfnamefont
  {R.}~\bibnamefont {Bratschitsch}},\ }\href@noop {} {\bibfield  {journal}
  {\bibinfo  {journal} {Nature Photonics}\ }\textbf {\bibinfo {volume} {2}},\
  \bibinfo {pages} {230} (\bibinfo {year} {2008})}\BibitemShut {NoStop}%
\bibitem [{\citenamefont {S{\"o}nnichsen}\ \emph {et~al.}(2002)\citenamefont
  {S{\"o}nnichsen}, \citenamefont {Franzl}, \citenamefont {Wilk}, \citenamefont
  {von Plessen}, \citenamefont {Feldmann}, \citenamefont {Wilson},\ and\
  \citenamefont {Mulvaney}}]{sonnichsen2002drastic}%
  \BibitemOpen
  \bibfield  {author} {\bibinfo {author} {\bibfnamefont {C.}~\bibnamefont
  {S{\"o}nnichsen}}, \bibinfo {author} {\bibfnamefont {T.}~\bibnamefont
  {Franzl}}, \bibinfo {author} {\bibfnamefont {T.}~\bibnamefont {Wilk}},
  \bibinfo {author} {\bibfnamefont {G.}~\bibnamefont {von Plessen}}, \bibinfo
  {author} {\bibfnamefont {J.}~\bibnamefont {Feldmann}}, \bibinfo {author}
  {\bibfnamefont {O.}~\bibnamefont {Wilson}}, \ and\ \bibinfo {author}
  {\bibfnamefont {P.}~\bibnamefont {Mulvaney}},\ }\href@noop {} {\bibfield
  {journal} {\bibinfo  {journal} {Physical Review Letters}\ }\textbf {\bibinfo
  {volume} {88}},\ \bibinfo {pages} {77402} (\bibinfo {year}
  {2002})}\BibitemShut {NoStop}%
\bibitem [{\citenamefont {Wang}\ and\ \citenamefont
  {Shen}(2006)}]{wang2006general}%
  \BibitemOpen
  \bibfield  {author} {\bibinfo {author} {\bibfnamefont {F.}~\bibnamefont
  {Wang}}\ and\ \bibinfo {author} {\bibfnamefont {Y.}~\bibnamefont {Shen}},\
  }\href@noop {} {\bibfield  {journal} {\bibinfo  {journal} {Physical Review
  Letters}\ }\textbf {\bibinfo {volume} {97}},\ \bibinfo {pages} {206806}
  (\bibinfo {year} {2006})}\BibitemShut {NoStop}%
\bibitem [{\citenamefont {Rivas}\ \emph {et~al.}(2004)\citenamefont {Rivas},
  \citenamefont {Kuttge}, \citenamefont {Bolivar}, \citenamefont {Kurz},\ and\
  \citenamefont {S{\'a}nchez-Gil}}]{rivas2004propagation}%
  \BibitemOpen
  \bibfield  {author} {\bibinfo {author} {\bibfnamefont {J.}~\bibnamefont
  {Rivas}}, \bibinfo {author} {\bibfnamefont {M.}~\bibnamefont {Kuttge}},
  \bibinfo {author} {\bibfnamefont {P.}~\bibnamefont {Bolivar}}, \bibinfo
  {author} {\bibfnamefont {H.}~\bibnamefont {Kurz}}, \ and\ \bibinfo {author}
  {\bibfnamefont {J.}~\bibnamefont {S{\'a}nchez-Gil}},\ }\href@noop {}
  {\bibfield  {journal} {\bibinfo  {journal} {Physical Review Letters}\
  }\textbf {\bibinfo {volume} {93}},\ \bibinfo {pages} {256804} (\bibinfo
  {year} {2004})}\BibitemShut {NoStop}%
\bibitem [{\citenamefont {Hibbins}\ \emph {et~al.}(2005)\citenamefont
  {Hibbins}, \citenamefont {Evans},\ and\ \citenamefont
  {Sambles}}]{hibbins2005experimental}%
  \BibitemOpen
  \bibfield  {author} {\bibinfo {author} {\bibfnamefont {A.~P.}\ \bibnamefont
  {Hibbins}}, \bibinfo {author} {\bibfnamefont {B.~R.}\ \bibnamefont {Evans}},
  \ and\ \bibinfo {author} {\bibfnamefont {J.~R.}\ \bibnamefont {Sambles}},\
  }\href@noop {} {\bibfield  {journal} {\bibinfo  {journal} {Science}\ }\textbf
  {\bibinfo {volume} {308}},\ \bibinfo {pages} {670} (\bibinfo {year}
  {2005})}\BibitemShut {NoStop}%
\bibitem [{\citenamefont {Lim}\ \emph {et~al.}(2007)\citenamefont {Lim},
  \citenamefont {Mar}, \citenamefont {Matheu}, \citenamefont {Derkacs},\ and\
  \citenamefont {Yu}}]{lim2007photocurrent}%
  \BibitemOpen
  \bibfield  {author} {\bibinfo {author} {\bibfnamefont {S.}~\bibnamefont
  {Lim}}, \bibinfo {author} {\bibfnamefont {W.}~\bibnamefont {Mar}}, \bibinfo
  {author} {\bibfnamefont {P.}~\bibnamefont {Matheu}}, \bibinfo {author}
  {\bibfnamefont {D.}~\bibnamefont {Derkacs}}, \ and\ \bibinfo {author}
  {\bibfnamefont {E.}~\bibnamefont {Yu}},\ }\href@noop {} {\bibfield  {journal}
  {\bibinfo  {journal} {Journal of Applied Physics}\ }\textbf {\bibinfo
  {volume} {101}},\ \bibinfo {pages} {104309} (\bibinfo {year}
  {2007})}\BibitemShut {NoStop}%
\bibitem [{\citenamefont {Catchpole}\ and\ \citenamefont
  {Polman}(2008)}]{catchpole2008plasmonic}%
  \BibitemOpen
  \bibfield  {author} {\bibinfo {author} {\bibfnamefont {K.}~\bibnamefont
  {Catchpole}}\ and\ \bibinfo {author} {\bibfnamefont {A.}~\bibnamefont
  {Polman}},\ }\href@noop {} {\bibfield  {journal} {\bibinfo  {journal} {Optics
  Express}\ }\textbf {\bibinfo {volume} {16}},\ \bibinfo {pages} {21793}
  (\bibinfo {year} {2008})}\BibitemShut {NoStop}%
\bibitem [{\citenamefont {Atwater}\ and\ \citenamefont
  {Polman}(2010)}]{atwater2010plasmonics}%
  \BibitemOpen
  \bibfield  {author} {\bibinfo {author} {\bibfnamefont {H.~A.}\ \bibnamefont
  {Atwater}}\ and\ \bibinfo {author} {\bibfnamefont {A.}~\bibnamefont
  {Polman}},\ }\href@noop {} {\bibfield  {journal} {\bibinfo  {journal} {Nature
  Materials}\ }\textbf {\bibinfo {volume} {9}},\ \bibinfo {pages} {205}
  (\bibinfo {year} {2010})}\BibitemShut {NoStop}%
\bibitem [{\citenamefont {Chen}\ \emph {et~al.}(2012)\citenamefont {Chen},
  \citenamefont {Agio},\ and\ \citenamefont
  {Sandoghdar}}]{chen2012metallodielectric}%
  \BibitemOpen
  \bibfield  {author} {\bibinfo {author} {\bibfnamefont {X.-W.}\ \bibnamefont
  {Chen}}, \bibinfo {author} {\bibfnamefont {M.}~\bibnamefont {Agio}}, \ and\
  \bibinfo {author} {\bibfnamefont {V.}~\bibnamefont {Sandoghdar}},\
  }\href@noop {} {\bibfield  {journal} {\bibinfo  {journal} {Physical Review
  Letters}\ }\textbf {\bibinfo {volume} {108}},\ \bibinfo {pages} {233001}
  (\bibinfo {year} {2012})}\BibitemShut {NoStop}%
\bibitem [{\citenamefont {Quinten}\ \emph {et~al.}(1998)\citenamefont
  {Quinten}, \citenamefont {Leitner}, \citenamefont {Krenn},\ and\
  \citenamefont {Aussenegg}}]{quinten1998electromagnetic}%
  \BibitemOpen
  \bibfield  {author} {\bibinfo {author} {\bibfnamefont {M.}~\bibnamefont
  {Quinten}}, \bibinfo {author} {\bibfnamefont {A.}~\bibnamefont {Leitner}},
  \bibinfo {author} {\bibfnamefont {J.}~\bibnamefont {Krenn}}, \ and\ \bibinfo
  {author} {\bibfnamefont {F.}~\bibnamefont {Aussenegg}},\ }\href@noop {}
  {\bibfield  {journal} {\bibinfo  {journal} {Optics Letters}\ }\textbf
  {\bibinfo {volume} {23}},\ \bibinfo {pages} {1331} (\bibinfo {year}
  {1998})}\BibitemShut {NoStop}%
\bibitem [{\citenamefont {Weeber}\ \emph {et~al.}(1999)\citenamefont {Weeber},
  \citenamefont {Dereux}, \citenamefont {Girard}, \citenamefont {Krenn},\ and\
  \citenamefont {Goudonnet}}]{weeber1999plasmon}%
  \BibitemOpen
  \bibfield  {author} {\bibinfo {author} {\bibfnamefont {J.-C.}\ \bibnamefont
  {Weeber}}, \bibinfo {author} {\bibfnamefont {A.}~\bibnamefont {Dereux}},
  \bibinfo {author} {\bibfnamefont {C.}~\bibnamefont {Girard}}, \bibinfo
  {author} {\bibfnamefont {J.~R.}\ \bibnamefont {Krenn}}, \ and\ \bibinfo
  {author} {\bibfnamefont {J.-P.}\ \bibnamefont {Goudonnet}},\ }\href@noop {}
  {\bibfield  {journal} {\bibinfo  {journal} {Physical Review B}\ }\textbf
  {\bibinfo {volume} {60}},\ \bibinfo {pages} {9061} (\bibinfo {year}
  {1999})}\BibitemShut {NoStop}%
\bibitem [{tes(2013)}]{test}%
  \BibitemOpen
  \href@noop {} {\enquote {\bibinfo {title} {Lumerical solutions inc.
  http://www.lumerical.com/tcad-products/fdtd/},}\ } (\bibinfo {year}
  {2013})\BibitemShut {NoStop}%
\bibitem [{\citenamefont {Shields}(2007)}]{shields2007semiconductor}%
  \BibitemOpen
  \bibfield  {author} {\bibinfo {author} {\bibfnamefont {A.~J.}\ \bibnamefont
  {Shields}},\ }\href@noop {} {\bibfield  {journal} {\bibinfo  {journal}
  {Nature Photonics}\ }\textbf {\bibinfo {volume} {1}},\ \bibinfo {pages} {215}
  (\bibinfo {year} {2007})}\BibitemShut {NoStop}%
\bibitem [{\citenamefont {M{\'a}rquez}\ \emph {et~al.}(2001)\citenamefont
  {M{\'a}rquez}, \citenamefont {Geelhaar},\ and\ \citenamefont
  {Jacobi}}]{marquez2001atomically}%
  \BibitemOpen
  \bibfield  {author} {\bibinfo {author} {\bibfnamefont {J.}~\bibnamefont
  {M{\'a}rquez}}, \bibinfo {author} {\bibfnamefont {L.}~\bibnamefont
  {Geelhaar}}, \ and\ \bibinfo {author} {\bibfnamefont {K.}~\bibnamefont
  {Jacobi}},\ }\href@noop {} {\bibfield  {journal} {\bibinfo  {journal}
  {Applied Physics Letters}\ }\textbf {\bibinfo {volume} {78}},\ \bibinfo
  {pages} {2309} (\bibinfo {year} {2001})}\BibitemShut {NoStop}%
\bibitem [{\citenamefont {Maier}(2007)}]{maier2007plasmonics}%
  \BibitemOpen
  \bibfield  {author} {\bibinfo {author} {\bibfnamefont {S.~A.}\ \bibnamefont
  {Maier}},\ }\href@noop {} {\emph {\bibinfo {title} {Plasmonics: Fundamentals
  and Applications}}}\ (\bibinfo  {publisher} {Springer},\ \bibinfo {year}
  {2007})\BibitemShut {NoStop}%
\bibitem [{\citenamefont {Rai-Choudhury}(1997)}]{rai1997handbook}%
  \BibitemOpen
  \bibfield  {author} {\bibinfo {author} {\bibfnamefont {P.}~\bibnamefont
  {Rai-Choudhury}},\ }\href@noop {} {\emph {\bibinfo {title} {Handbook of
  microlithography, micromachining, and microfabrication: microlithography}}},\
  Vol.~\bibinfo {volume} {1}\ (\bibinfo  {publisher} {SPIE Press},\ \bibinfo
  {year} {1997})\BibitemShut {NoStop}%
\bibitem [{\citenamefont {Blakemore}(1982)}]{blakemore1982semiconducting}%
  \BibitemOpen
  \bibfield  {author} {\bibinfo {author} {\bibfnamefont {J.}~\bibnamefont
  {Blakemore}},\ }\href@noop {} {\bibfield  {journal} {\bibinfo  {journal}
  {Journal of Applied Physics}\ }\textbf {\bibinfo {volume} {53}},\ \bibinfo
  {pages} {R123} (\bibinfo {year} {1982})}\BibitemShut {NoStop}%
\bibitem [{201(2014)}]{2014}%
  \BibitemOpen
  \href@noop {} {\enquote {\bibinfo {title} {Menzel gl{\"a}ser
  http://www.menzel.de/technical-information.656.0.html?\&l=1},}\ } (\bibinfo
  {year} {2014})\BibitemShut {NoStop}%
\bibitem [{\citenamefont {Davies}\ \emph {et~al.}(2012)\citenamefont {Davies},
  \citenamefont {Whittaker},\ and\ \citenamefont {Wilson}}]{davies2012metal}%
  \BibitemOpen
  \bibfield  {author} {\bibinfo {author} {\bibfnamefont {D.}~\bibnamefont
  {Davies}}, \bibinfo {author} {\bibfnamefont {D.}~\bibnamefont {Whittaker}}, \
  and\ \bibinfo {author} {\bibfnamefont {L.}~\bibnamefont {Wilson}},\
  }\href@noop {} {\bibfield  {journal} {\bibinfo  {journal} {Journal of Applied
  Physics}\ }\textbf {\bibinfo {volume} {112}},\ \bibinfo {pages} {044315}
  (\bibinfo {year} {2012})}\BibitemShut {NoStop}%
\bibitem [{\citenamefont {Jain}(2011)}]{Jain2011}%
  \BibitemOpen
  \bibfield  {author} {\bibinfo {author} {\bibfnamefont {P.}~\bibnamefont
  {Jain}},\ }\href@noop {} {\emph {\bibinfo {title} {Plasmonic Nanoparticles:
  Radiative and Non-Radiative Properties}}}\ (\bibinfo  {publisher} {VDM Verlag
  Dr. M{\"u}ller},\ \bibinfo {year} {2011})\BibitemShut {NoStop}%
\bibitem [{\citenamefont {Vollmer}\ and\ \citenamefont
  {Kreibig}(1995)}]{vollmer1995optical}%
  \BibitemOpen
  \bibfield  {author} {\bibinfo {author} {\bibfnamefont {M.}~\bibnamefont
  {Vollmer}}\ and\ \bibinfo {author} {\bibfnamefont {U.}~\bibnamefont
  {Kreibig}},\ }\href@noop {} {\emph {\bibinfo {title} {Optical Properties of
  Metal Clusters}}},\ Vol.~\bibinfo {volume} {25}\ (\bibinfo  {publisher}
  {Springer},\ \bibinfo {year} {1995})\BibitemShut {NoStop}%
\bibitem [{\citenamefont {Jackson}(1998)}]{Jackson1998}%
  \BibitemOpen
  \bibfield  {author} {\bibinfo {author} {\bibfnamefont {J.~D.}\ \bibnamefont
  {Jackson}},\ }\href@noop {} {\emph {\bibinfo {title} {Classical
  Electrodynamics}}}\ (\bibinfo  {publisher} {Wiley},\ \bibinfo {year}
  {1998})\BibitemShut {NoStop}%
\bibitem [{\citenamefont {Wokaun}\ \emph {et~al.}(1982)\citenamefont {Wokaun},
  \citenamefont {Gordon},\ and\ \citenamefont {Liao}}]{wokaun1982radiation}%
  \BibitemOpen
  \bibfield  {author} {\bibinfo {author} {\bibfnamefont {A.}~\bibnamefont
  {Wokaun}}, \bibinfo {author} {\bibfnamefont {J.}~\bibnamefont {Gordon}}, \
  and\ \bibinfo {author} {\bibfnamefont {P.}~\bibnamefont {Liao}},\ }\href@noop
  {} {\bibfield  {journal} {\bibinfo  {journal} {Physical Review Letters}\
  }\textbf {\bibinfo {volume} {48}},\ \bibinfo {pages} {957} (\bibinfo {year}
  {1982})}\BibitemShut {NoStop}%
\bibitem [{\citenamefont {Pfeiffer}\ \emph {et~al.}(2010)\citenamefont
  {Pfeiffer}, \citenamefont {Lindfors}, \citenamefont {Wolpert}, \citenamefont
  {Atkinson}, \citenamefont {Benyoucef}, \citenamefont {Rastelli},
  \citenamefont {Schmidt}, \citenamefont {Giessen},\ and\ \citenamefont
  {Lippitz}}]{doi:10.1021/nl102548t}%
  \BibitemOpen
  \bibfield  {author} {\bibinfo {author} {\bibfnamefont {M.}~\bibnamefont
  {Pfeiffer}}, \bibinfo {author} {\bibfnamefont {K.}~\bibnamefont {Lindfors}},
  \bibinfo {author} {\bibfnamefont {C.}~\bibnamefont {Wolpert}}, \bibinfo
  {author} {\bibfnamefont {P.}~\bibnamefont {Atkinson}}, \bibinfo {author}
  {\bibfnamefont {M.}~\bibnamefont {Benyoucef}}, \bibinfo {author}
  {\bibfnamefont {A.}~\bibnamefont {Rastelli}}, \bibinfo {author}
  {\bibfnamefont {O.~G.}\ \bibnamefont {Schmidt}}, \bibinfo {author}
  {\bibfnamefont {H.}~\bibnamefont {Giessen}}, \ and\ \bibinfo {author}
  {\bibfnamefont {M.}~\bibnamefont {Lippitz}},\ }\href@noop {} {\bibfield
  {journal} {\bibinfo  {journal} {Nano Letters}\ }\textbf {\bibinfo {volume}
  {10}},\ \bibinfo {pages} {4555} (\bibinfo {year} {2010})}\BibitemShut
  {NoStop}%
\bibitem [{\citenamefont {Bracher}\ \emph {et~al.}(2014)\citenamefont
  {Bracher}, \citenamefont {Schraml}, \citenamefont {Ossiander}, \citenamefont
  {Fr{\'e}d{\'e}rick}, \citenamefont {Finley},\ and\ \citenamefont
  {Kaniber}}]{bracher2014optical}%
  \BibitemOpen
  \bibfield  {author} {\bibinfo {author} {\bibfnamefont {G.}~\bibnamefont
  {Bracher}}, \bibinfo {author} {\bibfnamefont {K.}~\bibnamefont {Schraml}},
  \bibinfo {author} {\bibfnamefont {M.}~\bibnamefont {Ossiander}}, \bibinfo
  {author} {\bibfnamefont {S.}~\bibnamefont {Fr{\'e}d{\'e}rick}}, \bibinfo
  {author} {\bibfnamefont {J.}~\bibnamefont {Finley}}, \ and\ \bibinfo {author}
  {\bibfnamefont {M.}~\bibnamefont {Kaniber}},\ }\href@noop {} {\bibfield
  {journal} {\bibinfo  {journal} {Nanotechnology}\ }\textbf {\bibinfo {volume}
  {25}},\ \bibinfo {pages} {075203} (\bibinfo {year} {2014})}\BibitemShut
  {NoStop}%
\bibitem [{\citenamefont {Huang}\ \emph {et~al.}(2010)\citenamefont {Huang},
  \citenamefont {Callegari}, \citenamefont {Geisler}, \citenamefont
  {Br{\"u}ning}, \citenamefont {Kern}, \citenamefont {Prangsma}, \citenamefont
  {Wu}, \citenamefont {Feichtner}, \citenamefont {Ziegler}, \citenamefont
  {Weinmann} \emph {et~al.}}]{huang2010atomically}%
  \BibitemOpen
  \bibfield  {author} {\bibinfo {author} {\bibfnamefont {J.-S.}\ \bibnamefont
  {Huang}}, \bibinfo {author} {\bibfnamefont {V.}~\bibnamefont {Callegari}},
  \bibinfo {author} {\bibfnamefont {P.}~\bibnamefont {Geisler}}, \bibinfo
  {author} {\bibfnamefont {C.}~\bibnamefont {Br{\"u}ning}}, \bibinfo {author}
  {\bibfnamefont {J.}~\bibnamefont {Kern}}, \bibinfo {author} {\bibfnamefont
  {J.~C.}\ \bibnamefont {Prangsma}}, \bibinfo {author} {\bibfnamefont
  {X.}~\bibnamefont {Wu}}, \bibinfo {author} {\bibfnamefont {T.}~\bibnamefont
  {Feichtner}}, \bibinfo {author} {\bibfnamefont {J.}~\bibnamefont {Ziegler}},
  \bibinfo {author} {\bibfnamefont {P.}~\bibnamefont {Weinmann}},  \emph
  {et~al.},\ }\href@noop {} {\bibfield  {journal} {\bibinfo  {journal} {Nature
  Communications}\ }\textbf {\bibinfo {volume} {1}},\ \bibinfo {pages} {150}
  (\bibinfo {year} {2010})}\BibitemShut {NoStop}%
\bibitem [{\citenamefont {Giannini}\ \emph {et~al.}(2010)\citenamefont
  {Giannini}, \citenamefont {Berrier}, \citenamefont {Maier}, \citenamefont
  {S{\'a}nchez-Gil},\ and\ \citenamefont {Rivas}}]{giannini2010scattering}%
  \BibitemOpen
  \bibfield  {author} {\bibinfo {author} {\bibfnamefont {V.}~\bibnamefont
  {Giannini}}, \bibinfo {author} {\bibfnamefont {A.}~\bibnamefont {Berrier}},
  \bibinfo {author} {\bibfnamefont {S.~A.}\ \bibnamefont {Maier}}, \bibinfo
  {author} {\bibfnamefont {J.~A.}\ \bibnamefont {S{\'a}nchez-Gil}}, \ and\
  \bibinfo {author} {\bibfnamefont {J.~G.}\ \bibnamefont {Rivas}},\ }\href@noop
  {} {\bibfield  {journal} {\bibinfo  {journal} {Optics Express}\ }\textbf
  {\bibinfo {volume} {18}},\ \bibinfo {pages} {2797} (\bibinfo {year}
  {2010})}\BibitemShut {NoStop}%
\bibitem [{\citenamefont {Raithel}\ \emph {et~al.}(1997)\citenamefont
  {Raithel}, \citenamefont {Birkl}, \citenamefont {Phillips},\ and\
  \citenamefont {Rolston}}]{raithel1997compression}%
  \BibitemOpen
  \bibfield  {author} {\bibinfo {author} {\bibfnamefont {G.}~\bibnamefont
  {Raithel}}, \bibinfo {author} {\bibfnamefont {G.}~\bibnamefont {Birkl}},
  \bibinfo {author} {\bibfnamefont {W.}~\bibnamefont {Phillips}}, \ and\
  \bibinfo {author} {\bibfnamefont {S.}~\bibnamefont {Rolston}},\ }\href@noop
  {} {\bibfield  {journal} {\bibinfo  {journal} {Physical Review Letters}\
  }\textbf {\bibinfo {volume} {78}},\ \bibinfo {pages} {2928} (\bibinfo {year}
  {1997})}\BibitemShut {NoStop}%
\end{thebibliography}
\end{document}